\documentclass[conference]{IEEEtran}
\IEEEoverridecommandlockouts
\usepackage{cite}
\usepackage{amsmath,amssymb,amsfonts}
\usepackage{algorithmic}
\usepackage{graphicx}
\usepackage{textcomp}
\usepackage{xcolor}
\usepackage{array}
\usepackage{arydshln}
\usepackage{booktabs}
\usepackage{cite}
\usepackage{hyperref}
\usepackage{url}
\usepackage{comment}
\hypersetup{draft}

\begin{document}
\IEEEpubid{\makebox[\columnwidth]{979-8-3503-6491-0/24/\$31.00~\copyright2024~IEEE \hfill} \hspace{\columnsep}\makebox[\columnwidth]{ }}

\title{Ain't How You Deploy: An Analysis of BGP Security Policies Performance Against Various Attack Scenarios with Differing Deployment Strategies
}

\author{
\IEEEauthorblockN{
\begin{minipage}[t]{0.3\textwidth}
\centering
Seth Barrett\\
\textit{School of Computer and Cyber Sciences} \\
\textit{Augusta University}\\
Augusta, USA \\
sebarrett@augusta.edu
\end{minipage}
\hfill
\begin{minipage}[t]{0.3\textwidth}
\centering
Calvin Idom\\
\textit{College of Engineering and Science} \\
\textit{Louisiana Tech University}\\
Ruston, USA \\
cli025@email.latech.edu
\end{minipage}
\hfill
\begin{minipage}[t]{0.3\textwidth}
\centering
German Zavala Villafuerte\\
\textit{School of Computer Science} \\
\textit{Texas A\&M University - San Antonio}\\
San Antonio, USA \\
gzava010@jaguar.tamu.edu
\end{minipage}
}
\\
\IEEEauthorblockN{
\begin{minipage}[t]{0.45\textwidth}
\centering
Andrew Byers\\
\textit{School of Computing and Augmented Intelligence} \\
\textit{Arizona State University}\\
Tempe, USA\\
adbyers1@asu.edu
\end{minipage}
\hfill
\begin{minipage}[t]{0.45\textwidth}
\centering
Berk Gulmezoglu\\
\textit{Electrical and Computer Engineering} \\
\textit{Iowa State University}\\
Ames, USA \\
bgulmez@iastate.edu
\end{minipage}
}
}

\maketitle

\IEEEpubidadjcol

\begin{abstract} 
This paper investigates the performance of various Border Gateway Protocol (BGP) security policies against multiple attack scenarios using different deployment strategies. Through extensive simulations, we evaluate the effectiveness of defensive mechanisms such as Root Origin Validation (ROV), Autonomous System Provider Authorization (ASPA), and PeerROV across distinct AS deployment types. Our findings reveal critical insights into the strengths and limitations of current BGP security measures, providing guidance for future policy development and implementation.
\end{abstract}


\section{Introduction} 
\label{sec:introduction}

The Border Gateway Protocol (BGP) forms the backbone of the internet, routing information between autonomous systems (AS) and facilitating data exchange across vast networks. However, BGP's inherent lack of security measures makes it a frequent target for various cyber-attacks, including prefix hijacking and route leaks. These vulnerabilities not only compromise the integrity of data transmission but can also disrupt global communications. Existing defensive strategies, while beneficial, are often limited by partial deployment and lack comprehensive effectiveness across diverse network scenarios. This paper explores the performance of various BGP security policies through extensive simulations, providing insights into their effectiveness against different attack vectors under varied deployment strategies. We aim to highlight the strengths and weaknesses of current measures and propose robust solutions to enhance the security and resilience of BGP.

\section{Background} 
\label{sec:background}
The evolution of the BGP, the de facto system for routing information between autonomous systems on the internet, has been marked by persistent security challenges. Despite being critical for global internet infrastructure, BGP's original design lacks sufficient safeguards against malicious activities, leading to significant security vulnerabilities. Comprehensive surveys by Mitseva \textit{et al.} \cite{Survey_StateOfAffairs} and Huston \textit{et al.}. \cite{Survey_SecuringBGP} outline the extent of these vulnerabilities, ranging from prefix hijacking to route leaks. Moreover, efforts to secure BGP have been chronicled by Siddiqui \textit{et al.} \cite{Survey_RecentEfforts}, highlighting recent standardization attempts to mitigate these risks. These surveys collectively underscore the urgency of adopting robust security mechanisms, such as Resource Public Key Infrastructure (RPKI)\cite{nistNISTRPKI} and the Autonomous System Provider Authorization (ASPA)\cite{ASPA}, to strengthen BGP's resilience against enduring threats. The ongoing research and development in this domain are crucial for the implementation of effective countermeasures, as documented in the literature.

\section{Related Work}
\label{sec:relatedwork}

Previous studies regarding BGP security have mostly focused on detailed explanations of singular security aspects, or only a few. For example, the most comparable work in scope to our proposed idea is a survey on recent efforts in BGP security by Mitseva \textit{et al.} \cite{Survey_StateOfAffairs}, which outlines and gives a comprehensive review of all recent known attacks and relevant security measures, seeking to deal with these attacks. While it does encompass a large range of different kinds of attacks, it does not test them within any sort of simulation or environment in order to provide data to back up these claims. It contains real-world examples but does not take advantage of the versatility that a simulation environment can provide. Even more prevalent is the fact that this survey was completed in 2018. It has no mention of ASPA\cite{ASPA}, which is the newest and arguably most relevant innovation within BGP security, and it only has brief mentions of ROV and its variations\cite{Revisiting}, which have become the new cutting edge for the future of BGP security and are generally considered to have high potential.

In regards to ASPA works, the paper on Analysis of ASPA by Umeda \textit{et al.}\cite{Umeda_2023} is a more recent study, but still, it is only an analysis of ASPA's performance and does not compare it at a grander scale to other forms of security. Even though this work gives some insights into ASPA and utilizes LOTUS as a simulation tool in order to create quantitative data, its lack of comparison makes it challenging to relate the findings with BGP security. It doesn't provide a direction for security to move whilst referencing our results. Our study seeks to take this route to provide a comprehensive plan of action based on quantitative data that is recent and relevant. When these methods are evaluated in a simulation tool, it will recommend more realistic directions for the BGP community to develop new policies and implementations with a quantitative approach, as opposed to a comprehensive qualitative approach. 

Aforementioned reasons lead us into BGPy\cite{Furuness_2023} that provides a simulation framework for our experiments and the quantitative data for a better security analysis. BGPy is a simulation framework that allows users to run different kinds of attacks leveraging the latest CAIDA datasets\cite{CAIDA}, and evaluate how effective a chosen defense mechanism is toward the chosen attack. This laid the foundation of our empirical study, as many studies on BGP security has no empirical studies over the efficiency of available BGP defense mechanisms. For this reason, this study fills this gap in the research literature, and based on the results, recommend potential changes to further enhance the sustainability of BGP. 

\section{Methodology}
\label{sec:methodology}
Our purpose is to perform an empirical research, which evaluates the efficiency of defense mechanisms against a various of prominent attacks targeting BGP. The empirical results allow us to improve the advancement of BGP security and lead to an impact on the selection of implemented security protocols in future designs.

\subsection{Simulation Setup}
\label{subsec:simulationSetup}
To evaluate the efficacy of different defensive strategies against BGP attacks, we ran a series of simulations using BGPy\cite{Furuness_2023}, a simulator created for BGP security evaluation. Two separate sessions were used to run the simulations on Apple M1 and M2 MacBook architectures.

\subsubsection{First Simulation Session}
\label{subsubsec:firstSimSession}
The first session was run using BGPy at commit \texttt{de44b54} in the \texttt{master} branch. The execution of simulations was made easier by a custom simulation runner script, which is accessible on our GitHub repository\cite{seth}. We experimented with several deployment types of ASes, attack scenarios, and defensive policy combinations. \texttt{ROV}, \texttt{ASPA}, \texttt{PeerROV}, and \texttt{AS-Cones} were the defensive policies, whereas \texttt{Accidental Route Leak}, \texttt{Prefix Hijacking}, \texttt{Subprefix Hijacking}, and \texttt{Forged-Origin Prefix Hijacking} were the attack scenarios. There were 64 possible permutations due to the AS deployment types, which were \texttt{Input Clique}, \texttt{Stubs}, \texttt{Multihomed}, and \texttt{No Deployment Type}. There were 1000 trials in each combination, split up across six policy adoption percentages. The Simulation Details subsection (\ref{subsec:simulationDetails}) dives into greater detail about the simulations, including the scenarios, policies, and AS deployment types.

\subsubsection{Second Simulation Session}
\label{subsubsec:secondSimSession}
The second simulation session was run off of a M2 MacBook. We had originally intended to incorporate the ROV++\cite{ROV++} defensive policies in our first simulation session. This posed major problems in when attempting to implement code into BGPy\cite{Furuness_2023}. After contacting the creators of the BGPy library, they gave us a test version on a different branch named \texttt{real\_v8\_no\_bgpisec} which since been merged into \texttt{master}. This version included the \texttt{ROVPPv1Lite}, \texttt{ROVPPv2Lite}, and \texttt{ROVPPV2ImprovedLite} policy files. As in the last session, we executed ROV++ simulations across all attacks and deployment types, excluding `Accidental Route Leak`; this attack-policy combination gave errors from the simulation software that we were unable to completely fix.

\subsection{Simulation Details}
\label{subsec:simulationDetails}

\subsubsection{Policies}
\label{subsubsec:policies}
We incorporated in our simulations a range of protective measures that were intended to solve various BGP security issues:
\begin{itemize}
    \item \textbf{ROV:} One of the key components of the BGPy simulator for assessing the potency of BGP security measures is ROV. In order to stop misrouting brought on by prefix hijacking, it verifies a route's origin AS against a trusted registry\cite{Hlaváček_2023, DISCO}. The FCC recently issued a directive mandating that the top 10 most connected American ASes adopt ROV, highlighting the technology's vital role in improving network security and impacting our attention to the Input Clique deployment strategy of ROV in our simulations\cite{FCC}.
    
    \item \textbf{ASPA:} As previously discussed, ASPA is a protocol that aims to avoid route leaks and forged-origin prefix hijacks by enabling ASes to publish cryptographic objects that define approved routing connections\cite{ASPA}. This improves the security of BGP. The authors of BGPy built-in an implementation of ASPA, and we have incorporated it in our simulations to assess its efficacy because of its importance in tackling these particular BGP vulnerabilities, especially in light of recent guidelines that support its use in highly linked networks\cite{rodday2024exploring}.
    
    \item \textbf{PeerROV:} PeerROV is an implementation variation of ROV that filters BGP announcements only on the basis of peer connections. Within the adaptable simulation environment of BGPy, it provides a specific, context-dependent validation method by discarding notifications that ROA deems invalid if they come from peers.
    
    \item \textbf{AS-Cones:} Compared to ASPA, AS-Cones is a less developed path plausibility method that uses RPKI to protect AS relationship data in an attempt to reduce route leaks. Customer cones, or groups of ASes under the hierarchical supervision of a senior AS, are the emphasis of AS-Cones as opposed to ASPA. Since fewer ASes are needed to participate in AS-Cones than in ASPA, each AS publishes information detailing its customer cone, making it easier to discover and mitigate route leaks. Although it is not fully supported in BGPy yet, our implementation, which was motivated by Rodday et al.\cite{rodday2024exploring}, employs \texttt{OnlyToCustomers} as its basic policy, corresponding with the intended usage of AS-Cones in a real-world situation. This implementation is not perfect, and we are looking forward to its inclusion in BGPy. We believe that including it in the future will improve simulation power and accuracy while illustrating the dynamics of AS relationships.
    
    \item \textbf{ROV++:} 
    ROV++ Versions 1 and 2\cite{ROV++} are security measures proposed by Morillo \textit{et. al.}, which is meant to extend the base ROV protection over a BGP fabric that only has partial implementation of ROV. The different versions of ROV++ are included in BGPy as \texttt{ROVPPv1Lite}, \texttt{ROVPPv2Lite} and \texttt{ROVPPV2ImprovedLite}. Version 1 attempts to drop announcements sent to hijack sub-prefixes. Version 2 builds off of version one by using the blackhole announcement to compete for the sub-prefix in question, hopefully protecting other ASes from falling for the same false announcement. There is also a Version 3 being proposed, but was not covered in this analysis. The three policies mentioned are \texttt{LITE} versions of the original, which essentially means they were designed to be strictly software that is almost as effective as their full counterparts and are much easier to implement.
\end{itemize}

\subsubsection{Scenarios}
\label{subsubsec:scenarios}
A range of attack scenarios were simulated in order to evaluate the resilience of each defensive strategy:
\begin{itemize}
    \item \textbf{Accidental Route Leak} involves the unintentional spread of routing announcements beyond their intended scope, frequently as a result of setup errors, which has an impact on traffic flow and network efficiency.
    
    \item \textbf{Prefix Hijacking} occurs when a malevolent entity wrongfully asserts ownership of IP address blocks, diverting traffic meant for the rightful owner to themselves instead.
    
    \item \textbf{Subprefix Hijacking} focuses form of prefix hijacking in which smaller, more focused IP blocks are declared fraudulently; because of their specificity, these blocks are frequently harder to find.
    
    \item \textbf{Forged-Origin Prefix Hijacking} involves deceiving routers about the real source of the route by the malicious broadcast of prefixes with a fabricated origin AS.
\end{itemize}

\subsubsection{Deployment Types}
\label{subsubsec:deployTypes}
Various AS deployment options were considered to assess how the general efficacy of policies is impacted by their acceptance with different types of ASes:
\begin{itemize} 
    \item \textbf{No Deployment Type:} Since no particular AS type is the focus of this deployment strategy, the policies are implemented consistently to all AS types in the network. This kind of deployment is the most extensive and wide-ranging deployment option in our simulations, affecting all 77,029 ASes in the CAIDA Serial-2 dataset\cite{CAIDA}. As seen by the adoption data in Table \ref{tab:adoption_rates}, this technique naturally targets a greater number of ASes than any specific deployment type. 
    
    \item \textbf{Input Clique:} Usually Internet Service Providers (ISPs) or backbone providers encompass this type of AS. The Input Clique deployment type focuses on a densely linked core set of ASes. Because of their important locations within network architecture, these ASes are distinguished by their noteworthy effect on routing decisions. Our dataset contains only 19 ASes that fall under the definition of an input clique. This deployment type is highly similar to the type of ASes that the FCC prefers to implement ROV in their new directive\cite{FCC}.
    
    \item \textbf{Multihomed:} ASes that link to many ISPs in order to enhance redundancy and perhaps optimize routing decisions are known as multihomed ASes. Although these ASes don't have clients, they stay in touch with several peers or providers, which improves their resilience and connectivity. With 37,614 multihomed ASes in our dataset, their importance in preserving stable and adaptable network infrastructures is highlighted.
    
    \item \textbf{Stubs:} ASes that don't forward traffic for other ASes are known as stubs. These ASes, which are found near the edge of the internet, connect to a restricted set of providers and are mainly responsible for handling incoming and outgoing traffic for their own networks. 27,398 stub ASes make up a sizable fraction of the network's edge in our dataset, where edge-level security dynamics can be greatly impacted by the adoption of security rules.
\end{itemize}

\subsection{Data Analysis}
\label{subsec:dataAnalysis}
We had substantial amount of data to analyze after the completion of these two simulation sessions. Combining the data from the sessions was the first stage in our study. To simplify the data while keeping the most important fields, we created a simple Julia script that concatenates all the CSV files from the BGPy output directory into a single CSV file. Among these columns were \texttt{scenario\_cls}(the attack scenario), \texttt{AdoptingPolicyCls} (the primary policy), \texttt{PolicyCls} (BGP for all), \texttt{BasePolicyCls} (base policy, varying depending on the policy; see the Policies subsubsection(\ref{subsubsec:policies})), \texttt{percent\_adopt}, \texttt{outcome} (with possible values of \texttt{ATTACKER\_SUCCESS}, \texttt{VICTIM\_SUCCESS}, or \texttt{DISCONNECTED}), \texttt{value} (percentage of ASes hijacked, disconnected or successfully connected), \texttt{yerr} (y-error for 90\% confidence interval for the outcome), and \texttt{deployment type}. \\

Our primary focus was on the \texttt{VICTIM\_SUCCESS} outcomes, as we considered both the \texttt{DISCONNECTED} and \texttt{ATTACKER\_SUCCESS} outcomes as failures of the policy. We handled this data by utilizing many Exploratory Data Analysis (EDA) approaches in a Julia Jupyter notebook, which is also located in our repository\cite{seth}. Among our analyses were: 

\begin{itemize}
    \item \textbf{Bar Graphs:} In order to display the victim success rate of each deployment type, irrespective of adoption percentage, we created bar graphs for each policy-scenario combination. This made it easier to determine which policies worked effectively in all deployment scenarios.
    
    \item \textbf{Scenario-specific Bar Graphs:} We produced bar graphs showing the victim success percentages for every policy and deployment style for every scenario. This made it possible for us to evaluate which rules worked best against different types of scenarios by displaying them side-by-side.
    
    \item \textbf{2D Heatmaps:} We produced heatmaps for each policy-scenario permutation, where the victim success rate is represented by heat, to illustrate the link between adoption rate and deployment type. This graphic made it easier for us to identify how the cross between different deployment methods and adoption rates affected and whether the policies were effective.

    \item \textbf{Correlation Heat Maps:} In order to graphically represent the association between victim success rate and adoption percentage, we created correlation heat maps for every scenario-policy permutation. These maps helped identify the ways in which the percentage of ASes that adopted the policy in each scenario affected the success of the policy.
    
    \item \textbf{Cross Bar Graphs:} These plots revealed the distribution of y-error for every deployment type across all possible scenario-policy combinations, offering valuable information on the accuracy of our data. 
    
    \item \textbf{Multi-line Graphs:} We displayed the victim success rate for each deployment type as a function of adoption percentage in each scenario-policy combination, enabling us to see trends over a range of policy adoption levels.
\end{itemize}

Furthermore, for every combination of policy, scenario, and deployment type, we produced summary statistics that included mean, median, standard deviation, maximum, and lowest values for y-error, victim success rate, and adoption percentage. Our visual evaluations were quanitively supported by these statistics. 

Our comprehension of how effectively each policy worked against various situations across deployment types and adoption percentages was improved by this statistical analysis and visualization. We developed preliminary theories concerning the performance of policies based on the knowledge gained from our literature review. The following section on Findings (Section \ref{sec:findings}) discusses how we tested these hypotheses against our empirical data and how our findings supported or refuted them, along with which particular visualizations and EDA techniques contributed to these insights.

\section{Findings}
\label{sec:findings}

\subsection{Findings  Overview:}
There are many key issues and details that we have discovered over the course of analyzing the various graphs based on our experiments. These key patterns we noticed was input-clique only being 0.025\% of total AS type still performed well, at some points better than other deployment types. The policy that consistently performed well is ASPA with ROV as a base. We also noted no deployment type tends to outperform the other deployment types, this could be due to encompassing nearly all the ASes in the network. In the following section we will go more into details over our hypotheses and what conclusion we came to after our experiments. 

\subsection{Hypothesis and Results:}
\begin{itemize}
    \item \textbf{Hypothesis 1:} Higher adoption percentage by any policy or deployment combo should lead to a greater success rate.\\ 
    This hypothesis was shown to be proven wrong as the higher the percentage of adoption increased the victim success rate stayed the same, with the exceptions being ASPA and ROV. ASPA can actually be seen decreasing in victim success rate as the adoption percentage goes up, while ROV and its variations increase. This can be seen in Figures 1 and 8. 

    \begin{figure}
        \centering
        \includegraphics[width=\columnwidth]{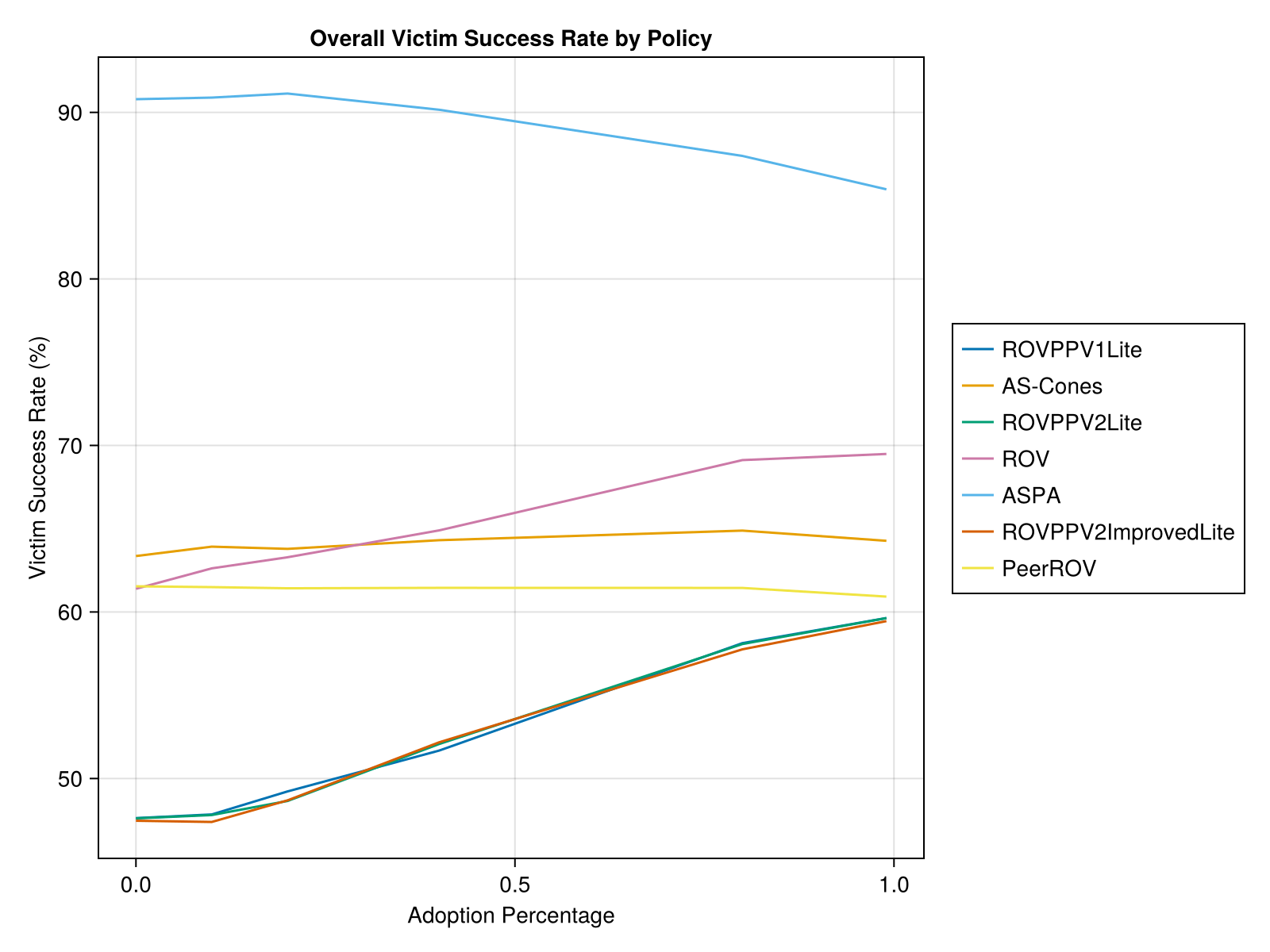}
        \caption{Multi-line graph representing the relationship between adoption percentage and victim success rate for each defensive policy}
    \end{figure}
    
    \item \textbf{Hypothesis 2:} ASPA should work best when deployed on stubs against any strategy at any adoption percent\cite{rodday2024exploring}.\\
    This hypothesis was proven to be mostly correct as ASPA was the most effective during a stub deployment type, although no deployment type did have a slight led at a lower adoption percentage. These results can be seen in Figures 2 and 3. Therefore, if one prefers to implement ASPA in a real world scenario it will be more efficient to deploy it at the stub, although the benefits are marginal. The cost of deployment would be a far more important factor to deployment type rather than the marginal differences in efficiency.
    \begin{figure}
        \centering
        \includegraphics[width=\columnwidth]{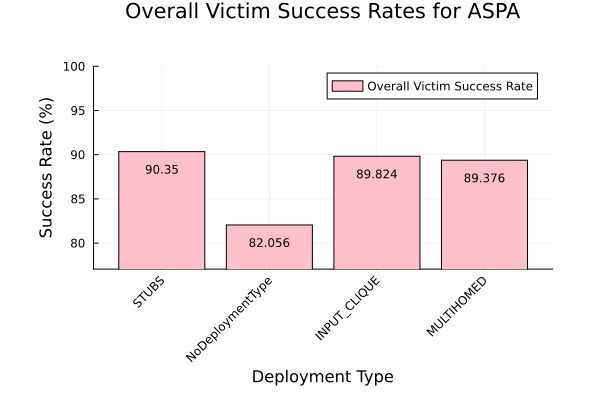}
        \caption{Bar graph representing the relationship between adoption percentage and victim success rate for each ASPA deployment strategy}
    \end{figure}
    \begin{figure}
        \centering
        \includegraphics[width=\columnwidth]{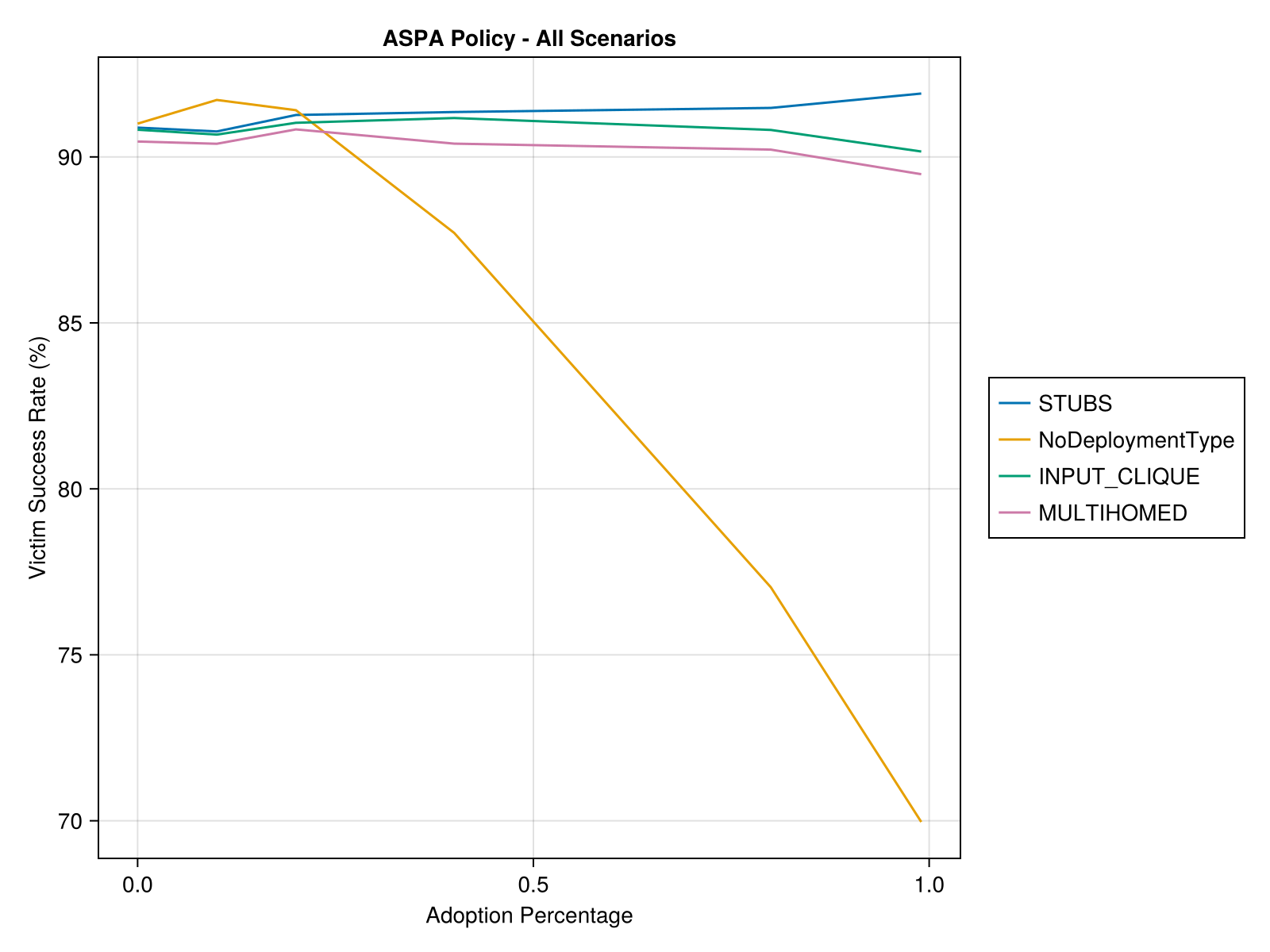}
        \caption{Multi-line graph representing the relationship between adoption percentage and victim success rate for each ASPA deployment strategy}
    \end{figure}

    \item \textbf{Hypothesis 3:} ROV will have the highest victim success rate compared to any policy against prefix and sub-prefix hijacking\cite{Umeda_2023,ASPA}.\\
    Based on Figures 4 and 5, it can be seen that ROV performs well in both prefix and sub-prefix but it is out-shadowed by ASPA. The reason is that since ASPA has ROV implementation as part of it with some additional policies and protections, which could cause the victim success rate to be higher. Additionally, enhanced ROV defence polices such as ROV++ performs better than ROV.
    \begin{figure}
        \centering
        \includegraphics[width=\columnwidth]{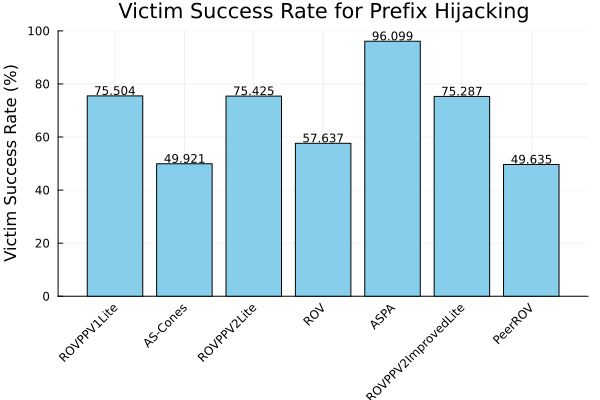}
        \caption{Bar graph representing victim success rate of defensive policies against Prefix Hijacking Attacks}
    \end{figure}
    \begin{figure}
        \centering
        \includegraphics[width=\columnwidth]{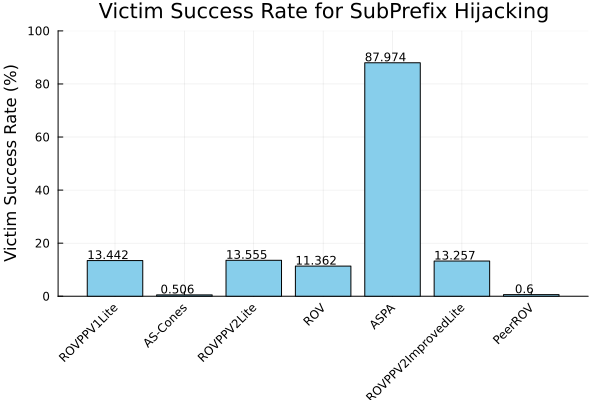}
        \caption{Bar graph representing victim success rate of defensive policies against Sub-Prefix Hijacking Attacks}
    \end{figure}
    

    \item \textbf{Hypothesis 4:} Any policy, especially ROV, will have a higher victim success rate as input-clique as the deployment compared to any other deployment type\cite{Furuness_2023, Survey_SecuringBGP, Survey_RecentEfforts, FCC}.\\ 
    By thoroughly examining the graphs in Figures 6 and 7, it can be observed that input-clique outperforms the other deployment types in certain combinations of attacks and defenses, but most of the time no deployment type is better. Although input-clique did not perform better than no deployment type, it would be far more simpler and cost effective deploying defenses to 19 ASes than nearly all of them, thus verifying the FCC's deployment strategy.
    \begin{figure}
        \centering
        \includegraphics[width=\columnwidth]{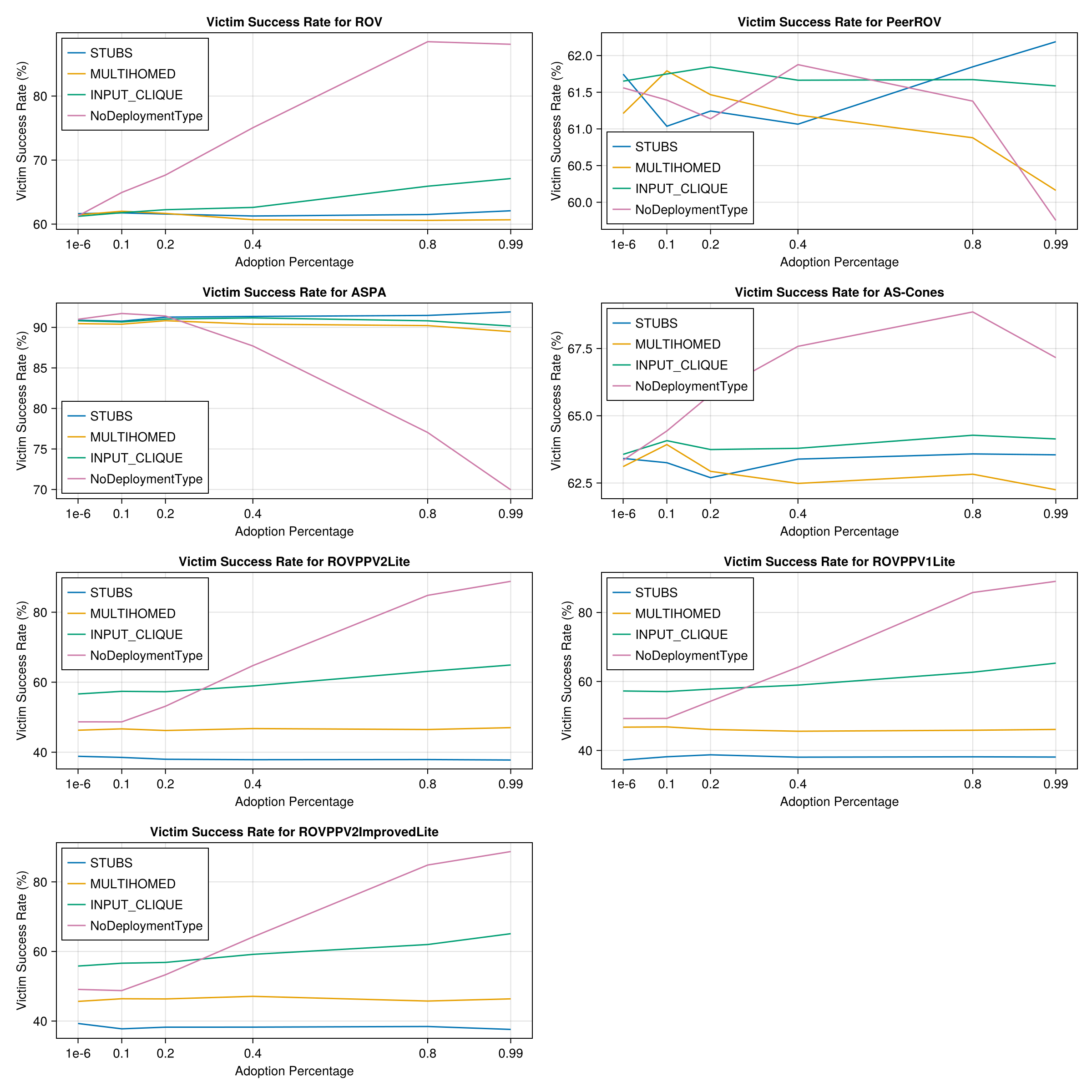}
        \caption{Multi-line Graphs representing victim success rates compared to adoption percentage for each deployment type and defensive policy}
    \end{figure}
    \begin{figure}
        \centering
        \includegraphics[width=\columnwidth]{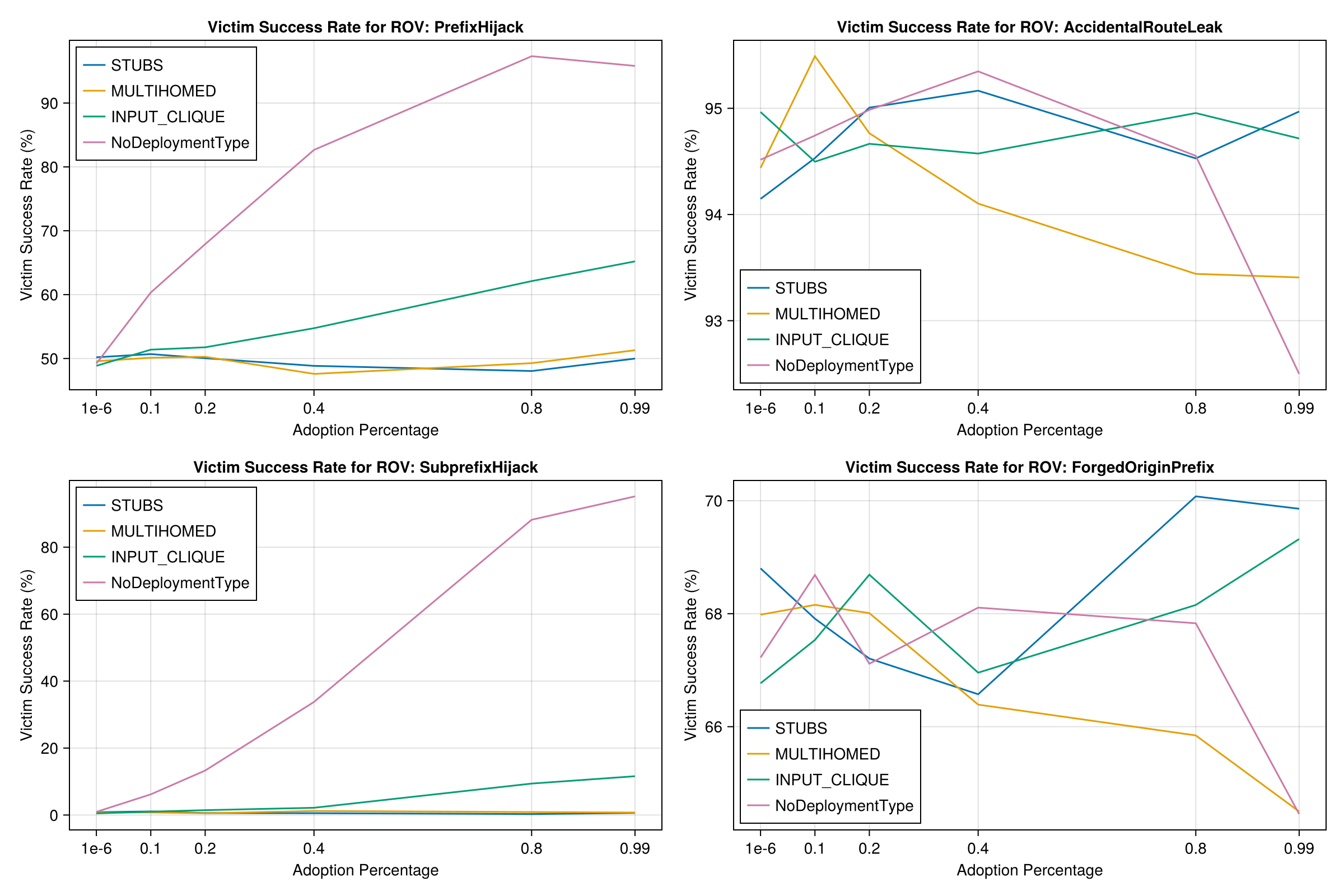}
        \caption{Multi-line graphs highlighting the comparison rate between adoption rate and victim success rate for each deployment type of ROV for each attack scenario}
    \end{figure}
\end{itemize}

\subsection{Issues}
\label{sec:issues}
We ran into several issues when conducting our study, including:
\begin{enumerate}
    \item \textbf{ROV++ Simulation:}
    One of the policies we originally set out to compare to was ROV++ and its different iterations (v1, v2, v3). Within the BGPy library, there were several useful pre-implemented policies that we were able to take advantage of in order to simulate security methods without having to implement it ourselves. This was effective for many methods, but not for ROV++. In its iteration on the website, it was unable to run, and its base heavily implemented very old code and imports that were no longer relevant in the current version, but were still dependencies. However, we reached out to the creators of BGPy\cite{Furuness_2023} and they responded very quickly saying they would be able to make an ROV++ policy simulation for us on short notice, but could not include v3 as it has potential security threats. We were only able to run it towards the end of our session time, and some of the data looked very similar across different versions of ROV++, but we were still able to get this data in for our final conclusion.

    \item \textbf{BGPy Multi-Platform Setup:}
    Since BGPy is a relatively new and small simulation tool, it was not as easy of a tool to get set up on our systems. We ran into several issues while trying to install it on all of our different systems due to differences in architecture and operating system from what BGPy was designed for. Even though our environments were set up correctly, attempting to run the program on our systems would be extremely challenging through several errors regarding each different dependency requiring a different version of Python, which had to be solved by manually opening each module and changing the required version in order to run. 

    \item \textbf{Simulation Runtime:}
    We set up our simulations to run 5 different levels of adoption percentages 200 times for a total of 1000 simulations for each possible combination of policy, scenario, and deployment type. In order to cover all possible variations of these 3 parameters, we had to run 64 simulations. Each of these simulations took approximately 15 minutes to run, for a total of 960 minutes (or 16 hours) for the simulations alone, even on our best system. Since we were able to leave it running on it's own, it wasn't terribly taxing, but running the simulation required all of the computer's power, rendering it useless to being able to perform any other tasks. Thankfully, we were able to get our simulations up and running rather early in the research cycle, but due to our brief time constraints, it could have potentially been a large problem if we ran into any other errors.

    \item \textbf{No Native Support for AS-Cones:}
    We stated in our original proposal that we wanted to include ASPA as a policy to analyze and consider in our grand conclusion, but one other approach that is essentially the same idea as ASPA is AS-Cones. AS-Cones are worth considering because we knew that ASPA would hold strong results from other papers, and AS-Cones are built on the same fundamentals as ASPA, which is creating AS objects that contain data about other authorized ASes. BGPy unfortunately did not have native support for this policy, but we believed this policy to be worth our time, so we instead implemented our own AS-Cones policy by referencing the structure defined by Rodday \textit{et al.} \cite{rodday2024exploring}

    \item \textbf{BGPy Output Could Use More Values:}
    Although BGPy has been an incredibly useful tool and has been the cornerstone for our empirical evaluations, we found that the data it produced in its CSV files could have provided more values, as the only values it gave us were the y-error value and the percentage of cases that received the type of propagation. Also of note, the creators of BGPy notified us that their \texttt{all but one} value does not work very well, so we had to use 99\% to get close enough.

    \item \textbf{Real-World Architecture Simulation:}
    During our study, we opted to simulate real-world scenarios rather than replicating the exact architecture used by ISPs. This decision was driven by the significant computing, time, and cost constraints associated with mimicking such complex infrastructures. While BGPy performed exceptionally well in its intended role as a simulation framework, enabling us to conduct our research without the need for high-performance computing (HPC), we recognize the potential value of conducting similar experiments on real-world architectures. We would be particularly interested in seeing an entity with the necessary resources undertake such a study, as it could provide deeper insights into BGP security mechanisms in practice. Additionally, the CAIDA dataset\cite{CAIDA} we used was from 2020, and while it included real-world information from ASes, utilizing a more up to date dataset might produce differing results.
\end{enumerate}

\section{Future Work}
\label{sec:futurework}
We plan to improve the implementation of ROV++ and AS-Cones, ensuring they work properly and are up-to-date. Additional simulations will focus on deploying these defenses to the top 10 ASes, evaluating the effectiveness of FCC's BGP regulation proposal. We also aim to deploy defenses to all ASes except one, rather than using 99\%. These efforts will provide valuable insights to enhance BGP security.

There are several sections we would like to improve on, such as the implementation of ROV++ and AS-Cones. Due to us implementing ROV++ based on technical explanation and a deadline, we would like to revisit it and perform some additional simulations to make sure it works properly. Similarly, for AS-Cones due to it still being in a developmental phase; there may be changes in the future we would need to into account for. Once we make sure these defense mechanisms are running properly and are up to date, we want to run a simulation where we only deploy the top 10 US-based ASes. The reason for this is to better determine the effectiveness of the proposal for BGP regulation created by FCC\cite{FCC}. This would allow us to see if the regulations would be enough to provide significant protection for BGP, or if a different direction would need to take place. Finally, we would want to deploy defenses to all ASes except one, rather than using 99\% in place of it. By running these simulations and obtaining more results, we hope to provide insight and knowledge to policy makers, researchers, and companies to better improve the security of BGP which it lacks. 

\section{Conclusion}
\label{sec:conclusion}
Our study identifies ASPA as the most effective BGP security mechanism, though broader deployment and further research are needed. PeerROV, despite its ease of implementation, offers limited protection and may not be the best priority for defense. Implementing defenses at the input clique level shows promise due to its effectiveness with fewer ASes involved, making it a practical and scalable solution. Future work would concentrate on cost-effective and practical strategies for deploying robust BGP security measures.

\section*{Acknowledgments}

We would like to acknowledge Neil Ziring from NSA for his proposal of the idea for this research. We also thank Iowa State University for hosting us onsite to work on this project. Additionally, Dr. Furuness, the creator of BGPy, was exceptionally helpful with his responses to our questions about BGPy. This material is based upon work supported by the National Science Foundation under Grant No. 1754048.. Additionally, this work was supported by grant H98230-21-1-0317.

\bibliographystyle{IEEEtran}
\bibliography{main}

\newpage
\onecolumn
\appendix

\begin{table}[h]
\centering

\begin{tabular}{|p{1.5cm}|p{.6cm}|c|c|c|c|c|}
\toprule
\textbf{Deployment Type} & \textbf{Only One} & \textbf{10\%} & \textbf{20\%} & \textbf{40\%} & \textbf{80\%} & \textbf{99\%} \\
\midrule          
Input Clique        & 1     & 2     & 4     & 8     & 16    & 19    \\
Stubs               & 1     & 2740  & 5480  & 10960 & 21919 & 27125 \\
Multihomed          & 1     & 3762  & 7523  & 15046 & 30092 & 37238 \\
No Deployment Type  & 1     & 7703  & 15406 & 30812 & 61624 & 76259 \\
\bottomrule
\end{tabular}
\vspace{.25cm}
\caption{Number of ASes Adopting Defensive Policy Based on Deployment Type and Adoption Percentages}
\label{tab:adoption_rates}
\end{table}

\vspace{2cm} 

\begin{figure}[h] 
        \centering
        \includegraphics[width=\textwidth]{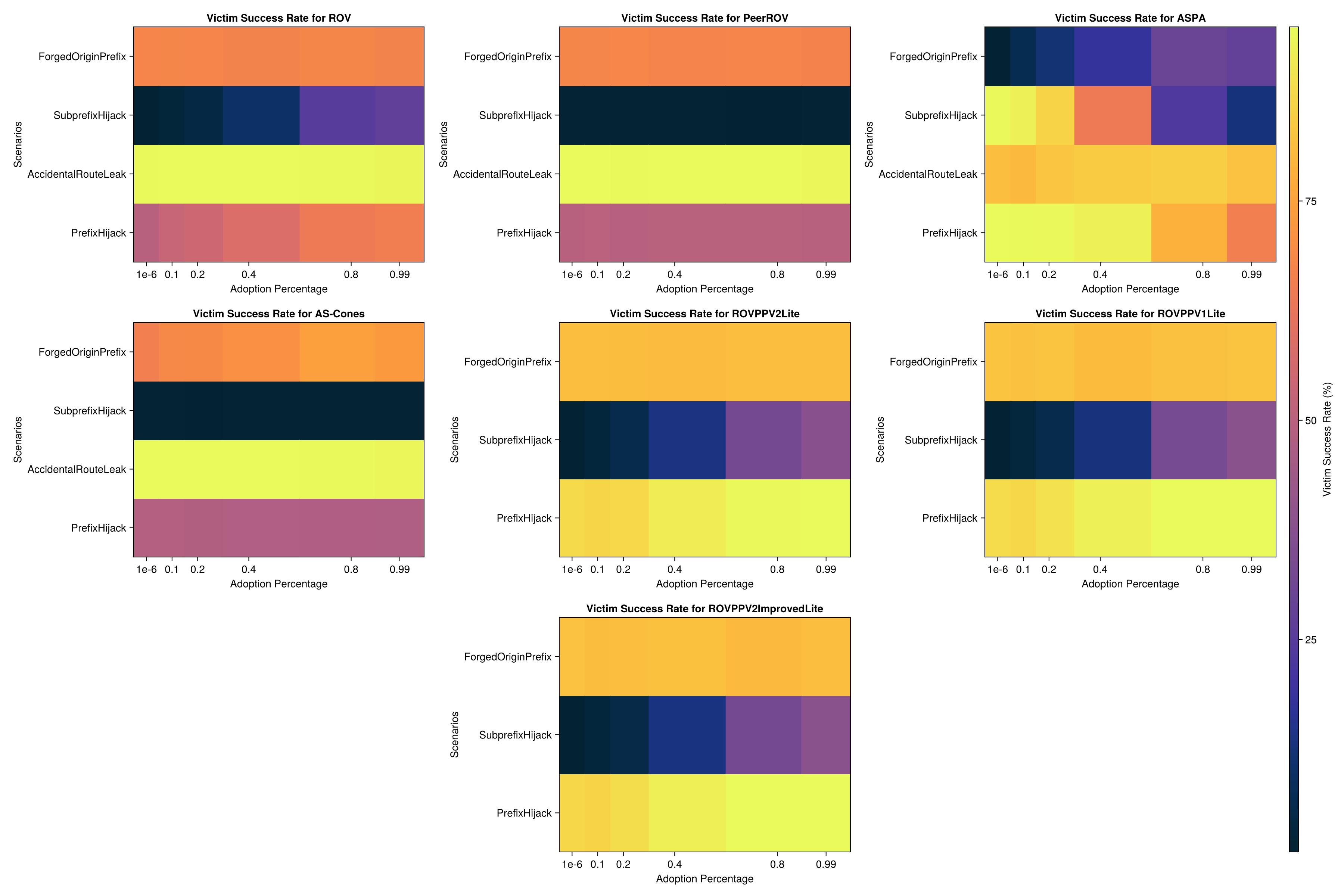}
        \caption{Heat-map illustrating the relationship between policy adoption rates and victim success rates across various attack scenarios.}
    \end{figure}

\end{document}